# A Robust Biometric-Based Three-factor Remote User Authentication Scheme

Vorugunti Chandra Sekhar
Infosys Technologies Limited
Electronics City, Bangalore, India
sekhar.daiict@gmail.com

Mrudula Sarvabhatla
S.V University
Tirupathi, India
mrudula.s911@gmail.com

## ABSTRACT
The rapid development of Internet of Things (IoT) technology, which is an inter connection of networks through an insecure public channel i.e. Internet demands for authenticating the remote user trying to access the secure network resources. In 2013, Ankita et al. proposed an improved three factor remote user authentication scheme. In this poster we will show that Ankita et al scheme is vulnerable to known session specific temporary information attack, on successfully performing the attack, the adversary can perform all other major cryptographic attacks. As a part of our contribution, we will propose an improved scheme which is resistance to all major cryptographic attacks and overcomes the defects in Ankita et al. scheme.

## Categories and Subject Descriptors
D.4.6, K.6.5

## General Terms
Security

## 1. INTRODUCTION
### 1.1 Registration Phase
R1. $U_i$ selects his identity $ID_i$, password $PW_i$ and computes W=h($PW_i$||N) and submits his biometrics information $B_i$ at fuzzy extractor to get Gen($B_i$) = ($R_i$,$P_i$), and submits $ID_i$, W to server S, through a secure channel, where N is a random number generated by $U_i$.

R2. After validating $ID_i$, S computes H=h($ID_i$||X), e = H$\oplus$W. S issues a user's smart card containing information {e, h(), p, g, Y} and sends it to $U_i$ through a secure channel. where Y = $g^X$ mod p.

R3. On receiving the smart card, $U_i$ computes L= N$\oplus R_i$ and V = h($ID_i$||$PW_i$||N). $U_i$ stores $P_i$, L, V into the smart card.

### 1.2 Login Phase
Whenever user $U_i$ wants to login to the remote server S, the user will perform the following steps.

L1. $U_i$ inserts his smart card into a card reader and inputs the $ID_i$, $PW_i$ and personal biometrics $B_i$ on fuzzy extractor to get $R_i$ = Rep($B_i$,$P_i$). On validating the $B_i$, smart card performs the following steps.

L2. Compute N = L$\oplus R_i$ and verify V* = h($ID_i$,$PW_i$,N) and check whether the computed V* equals to received V. if yes, S.C computes: H = e$\oplus$h($PW_i$||N), selects a random number $r_u$ and then computes : A1=$g^{ru}$ mod p, A2 = $Y^{ru}$ mod p = $(g^X)^{ru}$ mod p, NID = $ID_i \oplus$A2, $C_i$ = h($ID_i$||H||A1||A2||T1) and sends the login message <NID,A1,$C_i$,T1> to S, where T1 is the U's current timestamp.

### 1.3 Authentication Phase
After receiving the login request message, the remote server S will perform the following steps:

A1. On receiving the message < NID,A1,$C_i$,T1> at time T2, S verifies T2 - T1 $\leq \Delta$t, where $\Delta$t is the valid time delay. If verification does not hold, S terminates the session. Otherwise S computes A3 = $(g^{ru})^X$ mod p and retrieves $ID_i$ = NID$\oplus$A3. Then, it computes H = h($ID_i$||X) and verifies $C_i$ = h($ID_i$||H||A1||A3||T1), if $C_i$ equals the received $C_i$ then S proceeds further else terminates the session.

A2. S chooses a random number $r_s$ and computes A4 = $g^{rs}$ mod p, A5 = $(g^{ru})^{rs}$ mod p, Computes S.K = h($ID_i$||A3||A5||H||T1||T3), where T3 is the S current time stamp. S sends <Cs,A4,T3> to $U_i$ where Cs = h($ID_i$||S.K||H||T3). where T3 is the time at which 'S' sent the login reply message.

A3. On receiving the login message < Cs,A4,T3> at time T4, Smart card verifies T4-T3 $\leq \Delta$t. If yes , S.C computes A6 = $(g^{rs})^{ru}$ mod p, S.K = h($ID_i$||A2||A6||H||T1||T3). Finally it verifies, $C_s^*$ = h($ID_i$||S.K||H||T3). S.C verifies the $C_S^*$ = $C_S$. If verification doesn't hold, S.C terminates the session. On mutual authentication among $U_i$ and server S, all the further communication is encrypted with the session key framed.

## 2. Cryptanalysis of Ankita et al Scheme
To analyze the security of Ankits et al.'s scheme, we assume that an attacker could obtain the secret values stored in the smart card of $U_i$ i.e {$e_i$, h(), p, g, Y, $P_i$, L,V } by monitoring the power consumption [2] and the intercepted messages i.e <NID,A1,$C_i$,T1>, <Cs,A4,T3> between the user and the server.

### 2.1 Known Session Specific Temporary Information Attack
If an adversary 'E' gets the session secret value of user $U_i$. i.e $r_u$ and biometric information $B_i$, 'E' can perform following steps.

Step1: Frame A1=$g^{ru}$ mod p, A2 = $Y^{ru}$ mod p.

Step 2: 'E' can frame $ID_i$ = NID$\oplus$A2, from intercepted login message <NID,A1,$C_i$,T1>.

Step 3: Compute: Rep ($B_i$,$P_i$) = $R_i^*$, N = L$\oplus R_i^*$. Intercept H = e$\oplus$W and replace H with e$\oplus$W = e$\oplus$h($PW_i$||N) in $C_i$ equation. i.e $C_i$ = h($ID_i$||H||A1||A3||T1)

Step 4: Frame $C_i = h(ID_i \| e \oplus h(PW_i \| N) \| A1 \| A2 \| T1)$. In $C_i$, 'E' knows all the values of $U_i$ except $PW_i$. Now 'E' can perform password guessing attack on $C_i$. Guess the value of $PW_i$ to be $PW_i^*$ from uniformly distributed dictionary and check $C_i^* = h(ID_i \| e \oplus h(PW_i^* \| N) \| A1 \| A2 \| T1)$. If both sides are equal, then the $U_i$ password is $PW_i^*$. Otherwise 'E' can repeat the process to get correct value $PW_i^*$. On getting correct password $PW_i$, 'E' can frame $H = e \oplus h(PW_i^* \| N)$.

Step 5: Compute $A6 = (A4)^{ru} \mod p = (g^{rs})^{ru} \mod p$ from intercepted message $<Cs, A4, T3>$.

On getting $PW_i$, $ID_i$, N as discussed above, the adversary can frame the session key $S.K = h(ID_i \| A2 \| A6 \| H \| T1 \| T3)$ and can perform all major cryptographic attacks like user impersonation, DoS, Masquerade attacks etc.

## 3. PROPOSED SCHEME

### 3.1 Registration Phase
R1. $U_i$ selects his identity $ID_i$, password $PW_i$ and computes $W=h(PW_i \| N \| T1)$ and submits his biometrics information $B_i$ at fuzzy extractor to get $Gen(B_i) = (R_i, P_i)$, and submits $ID_i$, W to server S at time T1, through a secure channel, where N is a random number generated by $U_i$.

R2. On receiving the login message at time T2, After validating $ID_i$, S computes $G=h(ID_i \| X)$, $H=G \oplus T2$, $e = H \oplus W$. S issues a user's smart card containing information $\{e, h(), p, g, Y\}$, $T1 \oplus T2$ to $U_i$ through a secure channel. where $Y = g^X \mod p$.

R3. On receiving the smart card, $U_i$ computes $L= N \oplus R_i \oplus T1$ and $V = h(ID_i \| T1 \| PW_i \| T2 \| N)$. $U_i$ stores $P_i$, L, V, $M= h(ID_i \oplus T2) \oplus T1$, $N=h(PW_i \| R_i) \oplus T2$ into the smart card.

### 3.2 Login Phase
Whenever user $U_i$ wants to login to the remote server S, the user will perform the following steps.

L1. $U_i$ inserts his smart card into a card reader and inputs the $ID_i$, $PW_i$ and personal biometrics $B_i$ on fuzzy extractor to get $R_i = Rep(B_i, P_i)$. On validating the $B_i$, smart card performs the following steps.

L2. Compute $T2 = N \oplus h(PW_i \| R_i)$, $T1= M \oplus h(ID_i \oplus T2)$, $N = R_i \oplus L \oplus T1$ and verify $V^* = h(ID_i \| T1 \| PW_i \| T2 \| N)$ and check whether the computed $V^*$ equals to received V. if yes, S.C computes: $H = e \oplus h(PW_i \| N \| T1)$. selects a random number $r_u$ and then compute $A1=g^{ru} \mod p$, $A11 = A1 \oplus T2 \oplus T3$, $A2 = Y^{ru} \mod p = (g^X)^{ru} \mod p$, $A22 = A2 \oplus T3$, $NID = ID_i \oplus A22 \oplus h(T1 \| T3 \| T2)$, $C_i = h(ID_i \| H \| A22 \| A11 \| T1 \| T3 \| T2)$ and sends the login message $<NID, A11, C_i, Q=T3 \oplus h(T1)>$ to S where T3 is the current time of smart card.

### 3.3 Authentication Phase
A1. On receiving the message $< NID, A11, C_i, T3 \oplus h(T1)>$ at time T4, S verifies $T4 - T3 \leq \Delta t$. if yes, S computes $Q \oplus h(T1) = T3$. $A11 \oplus T2 \oplus T3 = A1^*$, $A2^* = (A1^*)^X \mod p$, $A22^* = A2^* \oplus T3$, and retrieves $ID_i = NID \oplus A22^* \oplus h(T1 \| T3 \| T2)$. Then, 'S' computes $H = h(ID_i \| X) \oplus T2$ and verifies $C_i^* = h(ID_i \| H \| A22^* \| A11^* \| T1 \| T3 \| T2)$, if $C_i^*$ equals the received $C_i$ then S proceeds further else terminates the session.

A2. S chooses a random number $r_s$ and computes $A4 = g^{rs} \mod p$, $A44 = A4 \oplus T3 \oplus T4$, $A5 = (g^{ru})^{rs} \mod p$, $A55 = A5 \oplus T3 \oplus T5$. Computes $S.K = h(ID_i \| A22^* \| A55 \| H \| T1 \| T3 \| T5)$, where T5 is the S current time stamp. S further computes $C_s = (ID_i \| S.K \| H \| T2 \| T4)$, $P=h(T1 \| ID_i \| T3) \oplus T4$, $Q=h(T2 \| ID_i \| T3) \oplus T5$, S sends $<Cs, A44, P, Q>$ to $U_i$ at time T5.

A3. On receiving the login message $<Cs, A44, P, Q>$ S.C computes $T4 = P \oplus h(T1 \| ID_i \| T3)$, $T5 = Q \oplus h(T2 \| ID_i \| T3)$, $A44 \oplus T3 \oplus T4 = A4$, $A5^* = (A4)^{ru} \mod p = (g^{rs})^{ru} \mod p$, $A55^* = A5^* \oplus T3 \oplus T5$, $S.K = h(ID_i \| A22 \| A55^* \| H \| T1 \| T3 \| T5)$. Finally S.C computes $C_S^* = h(ID_i \| S.K \| H \| T2 \| T4)$ and verifies whether $C_S^* = C_S$. If verification doesn't hold, S.C terminates the session. Once the $U_i$ and S are mutually authenticated, all the further communication is encrypted with the session key framed.

## 4. SECURITY ANALYSIS OF PROPOSED SCHEME

**Table 1. Type of users and the values they know**

| Types of User | Values known to the user | Values doesn't known to the user |
|---|---|---|
| Legal user ($U_i$): A user who is legitimated and trying to access the system with his own smart card. | A legal user knows his own $ID_i$, $PW_i$, $B_i$, e, h(), p, g, Y, $P_i$, L, V, M, N, $r_u$, T1,T2,T3, T4,T5 | X, $r_s$ |
| Legal Adversary (E)+ Insider : A user who is legitimated and trying to access the system with the stolen smart card of another legal user ($U_i$), having access to all messages exchanged between $U_i$, S. | 1. Access to e, h(), p, g, Y, $P_i$, L, V, M, N from the $U_i$ smart card. 2. $<NID, A11, C_i, Q=T3 \oplus h(T1)>$, $<Cs, A44, M, N>$ by intercepting the login messages exchanged among $U_i$, S. 3. $B_i$, $R_i$, $P_i$, $r_u$, $r_s$ (Assumed to be known to the adversary.) | $ID_i$, $PW_i$, T1, T2, T3, T4, T5, X, $r_s$ |

**Table 2. Cost comparison among various smart card schemes**

| Type of Cost | Proposed Scheme | Ankita et al. [1] |
|---|---|---|
| Total Computation Cost | $21T_{Hash}$ | $11T_{Hash}$ |
| Communication Overhead | 128*8 | 128*7 |
| Storage Cost | 128*10 | 128*8 |

$ID_i$, Time stamp: 128 bits. Output of Hash function is 128-bit.

In our scheme the legal adversary 'E' is assumed to know $B_i$, $r_u$, $r_s$, smart card values of legal user $U_i$. Due to no chance of getting the values $ID_i$, T1, T2, T3, T4 and T5 of $U_i$, it's not possible for 'E' to guess any unknown value of $U_i$ and to perform any kind of attack, where as in Ankita et al. scheme with $B_i$, $r_u$ value the adversary can come to know the passowrd of $U_i$ and can perform all major atatcks. Hence we conclue that with negligible increase in computation, communication and storage cost we have proposed a robust remote user authentication scheme which is resistant to all major attacks.